\documentclass[a4paper,11pt]{article}
\usepackage{jcappub}
\usepackage{graphicx}
\usepackage{subfig}
\usepackage{dcolumn}
\usepackage{bm}
\usepackage{hyperref}
\usepackage{braket}
\usepackage{color}
\usepackage{amsmath}
\usepackage{float}
\usepackage{orcidlink}
\usepackage{slashed}
\usepackage{ulem}

\usepackage{caption}
\usepackage{tikz,xcolor,hyperref}
\definecolor{darkgreen}{rgb}{0.0, 0.2, 0.13}
\definecolor{bostonuniversityred}{rgb}{0.8, 0.0, 0.0}
\definecolor{lime}{HTML}{A6CE39}

\newcommand{\SN}{\mathrm{SN}}
\newcommand{\obs}{\mathrm{obs}}
\newcommand{\DM}{\mathrm{DM}}

\newcommand{\rSN}{\overrightarrow{r_{\SN}}}
\newcommand{\rE}{\overrightarrow{r_{\oplus}}}
\newcommand{\xvec}{\overrightarrow{x}}

\title{The Dark Matter Diffused Supernova Neutrino Background}

\author[a,b]{Garv Chauhan\,\orcidlink{0000-0002-8129-8034}}
 \affiliation[a]{Department of Physics, Arizona State University, 450 E. Tyler Mall, Tempe, AZ 85287-1504 USA}
\affiliation[b]{Center for Neutrino Physics, Department of Physics, Virginia Tech, Blacksburg, VA 24061, USA}

\author[b,c]{R. Andrew Gustafson\,\orcidlink{
0000-0002-4794-7459}}
\affiliation[c]{International Center for Quantum-field Measurement Systems for Studies of the Universe and Particles (QUP,WPI), High Energy Accelerator Research Organization (KEK), Oho 1-1, Tsukuba, Ibaraki 305-081, Japan}

\author[b,d,e]{Gonzalo Herrera\,\orcidlink{0000-0001-9250-8597}}

\affiliation[d]{Department of Physics and Kavli Institute for Astrophysics and Space Research, Massachusetts Institute of Technology, Cambridge, MA 02139, USA}

\affiliation[e]{Harvard University, Department of Physics and Laboratory for Particle Physics and Cosmology, Cambridge, MA 02138, USA}

\author[b]{Taj Johnson}

\author[b]{Ian M. Shoemaker\,\orcidlink{
0000-0001-5434-3744}}

\abstract{We consider neutrinos scattering off Milky Way dark matter and the impact of this scattering on supernovae neutrinos. This can take the form of attenuation on the initial flux of neutrinos and a time-delayed flux of scattered neutrinos. Considering dark matter masses above 100 MeV and past Milky Way supernovae, we find this time-delayed flux is nearly constant in time. We call this flux the Dark Matter Diffused Supernova Neutrino Background (DMDSNB), and use Super-K limits on the Diffuse Supernova Neutrino Background (DSNB) flux to set limits on the dark matter-neutrino scattering cross section. We find $\sigma_{\DM-\nu}/m_{\rm DM} \lesssim 2.4 \times 10^{-24} \mathrm{cm^2}$/GeV for $m_{\rm DM} \gtrsim 1$ GeV, which is the strongest bound to date on dark matter-neutrino scatterings at MeV energies, and stronger than bounds set from SN1987A neutrino attenuation by an order of magnitude. We end by discussing how the DMDSNB could be distinguished from the DSNB.}

\begin{document}
\maketitle
\flushbottom

\section{Introduction}
Dark matter coupled solely to neutrinos can yield the observed relic abundance of the Universe in wide regions of parameter space yet untested experimentally, ranging from a dark matter mass of $m_{\rm DM} \sim 10$ keV to $m_{\rm DM} \sim 100$ TeV \cite{Lindner:2010rr, vandenAarssen:2012vpm,Dasgupta:2013zpn,Bertoni:2014mva,Cherry:2014xra,Olivares-DelCampo:2017feq, Berlin:2017ftj,Berlin:2018ztp,Blennow:2019fhy, Arguelles:2019ouk, Cardenas:2024ojd}. In contrast with models in which the dark matter couples to electromagnetically charged Standard Model (SM) particles, this scenario is not well-probed at direct detection experiments, owing to loop-suppressed couplings to electrons and nuclei\footnote{One exception is given by the scenario where the dark matter couples to the lepton doublet as a whole, resulting in a comparable coupling strength of dark matter to neutrinos and $\textit{e.g}$ electrons. In that situation, limits on the dark matter-neutrino cross section have been derived from the consideration that astrophysical neutrino fluxes can induce a flux of boosted dark matter at Earth, detectable at direct detection experiments via electron scattering \cite{Das:2021lcr, Lin:2023nsm, Ghosh:2021vkt, Zhang:2020nis, Sun:2025gyj, Ghosh:2024dqw, Lin:2024vzy, Das:2024ghw, DeRomeri:2023ytt}. While interesting, this scenario requires an additional assumption on the nature of dark matter microphysics, and introduces a degeneracy on the measurement among the two coupling strength. In this work we look at the modification of astrophysical neutrino fluxes due to interactions with dark matter during propagation in the Milky Way, which does not suffer from these caveats.}. However, phenomenological tests of a dark matter-neutrino interaction have been discussed in a number of other contexts, such as colliders and beam dump experiments \cite{Berryman:2022hds, Dev:2024twk}, cosmology \cite{Boehm:2000gq,Mangano:2006mp, Wilkinson:2014ksa, Escudero:2015yka, Boehm:2014vja, Brax:2023tvn}, and a number of astrophysical settings~\cite{Farzan:2014gza,Cherry:2014xra,Arguelles:2017atb,Kelly:2018tyg, Choi:2019ixb,Murase:2019xqi,GonzaloTXSAtten,Carpio:2022sml,Cline:2022qld,Eskenasy:2022aup,Fujiwara:2023lsv,Cline:2023tkp,GonzaloTDEDiffusion, KA:2023dyz, Tseng:2024akh, Heston:2024ljf}.

Of particular interest are the constraints derived on dark matter-neutrino interactions from astrophysical neutrino sources, which overcome cosmological bounds in models where the dark matter-neutrino scattering cross section rises with the incoming energy of the neutrino. Some astrophysical neutrino sources considered in this context are Active Galactic Nuclei \cite{GonzaloTXSAtten, Cline:2022qld, Cline:2022qld}, Tidal Disruption Events \cite{Fujiwara:2023lsv, GonzaloTDEDiffusion}, the Sun \cite{Capozzi:2017auw}, and supernovae. Focusing on supernovae, bounds have been derived from the non-observation of a significant attenuation in the neutrino flux from SN1987A due to scatterings with dark matter in the intergalactic medium and in the Milky Way \cite{Kolb:1987qy}, and from the absence of an anomalous energy loss induced by a dark matter coupling to neutrinos \cite{Cappiello:2025tws}. Further projected bounds have been proposed from the scatterings of diffuse supernova neutrinos off dark matter \cite{Farzan:2014gza}, and from future observations of
 nearby supernova neutrinos \cite{Murase:2019xqi,Carpio:2022sml}.

Here we propose a new phenomenological probe of dark matter-neutrino interactions using supernovae. We discuss the diffuse neutrino flux induced by dark matter scattering with neutrinos of galactic supernova origin. We dub this flux the \textit{Dark Matter Diffused Supernova Neutrino Background}. This is in contrast with the usual diffuse supernova neutrino background, which is produced from extragalactic supernovae \cite{Ando:2004hc, Horiuchi:2017qja}.  We consider the expected spatial distribution of galactic supernovae and their rate of occurrence, together with the dark matter distribution in the Galaxy to determine this flux. We discuss both the spectral and timing signatures of this flux, and derive novel constraints on the dark matter-neutrino scattering cross sections. We also revisit bounds on the dark matter-neutrino cross section from SN1987A via an attenuation of the neutrino fluxes, and derive bounds from the non-observation of a significant time-delayed neutrino flux from SN1987A.

\section{Attenuation and Time Delay}
Let us consider neutrinos from a supernova at $\rSN$, which will travel to Earth at $\rE$ (define $d = |\rSN - \rE|$ for convenience). We will now define the optical depth between these two points as

\begin{equation}
    \tau(\rSN,\rE) = \frac{\sigma_{\DM-\nu}}{m_{\DM}} \int_{\rSN}^{\rE} \rho_{\DM} (\overrightarrow{r})dl \label{eq-optical-depth},
\end{equation}
where $m_{\DM} $ is the dark matter mass, $\rho_{\DM}(\overrightarrow{r})$ is the dark matter energy density at a location $\overrightarrow{r}$, and $\sigma_{\DM-\nu}$ is the total dark matter-neutrino cross section. If we consider any time delays from scattering to be much larger than the timescale of neutrino production (which we will see later is a good assumption), then the attenuated fluence of neutrinos at Earth within the first $\sim 10$ seconds is

\begin{equation}
    \frac{dF_{a}}{dE_{\nu}} \bigg|_{\mathrm{Direct}} = \frac{\exp\big(-\tau(\rSN,\rE) \big)}{4 \pi d^2}\frac{dN_{\nu}}{dE_{\nu}},  \label{eq-atten-fluence}
\end{equation}
where $dN_{\nu}/dE_{\nu}$ is the differential production of neutrinos from the supernova.

Turning now to the idea of neutrino time delays, we will consider the dark matter-neutrino scattering to occur at $\xvec$. Let us define  $d_{x, \oplus}$, and $d_{x, \SN}$ to be, $|\rE - \xvec|$, and $|\rSN - \xvec|$ respectively. It is useful to redefine our expression into observable quantities, so we will use neutrino time delay (relative to the first neutrino arrival) $t = d_{x,\SN} + d_{x,\oplus}- d$, observed neutrino energy $E_{\obs}$, and observed solid angle $\Omega_{\obs}$ ($\theta_{\obs} = 0$ corresponds to the observed neutrino pointing directly back to the supernova). Putting this together, treating the supernova neutrino emission as an instantaneous event and accounting for the energy distribution of the source, we find an expression for the observed time-delayed flux.

\begin{equation}
\begin{split}
    \frac{d \Phi_{\obs}}{dE_{\obs} d\Omega} =& \int d^{3}\xvec dE_{\nu} \frac{dN}{dE_{\nu}}  \bigg(\frac{\exp(-\tau(\rSN,\xvec)}{4 \pi d_{x,\SN}^2} \bigg) \bigg( \frac{\rho_{\DM}(\xvec)}{m_{\DM}} \frac{\exp(-\tau(\xvec,\rE)}{ d_{x,\oplus}^2} \frac{d \sigma_{\DM-\nu}}{d \Omega_{s}}\bigg)\\ & \times \delta(t-t(\xvec))
    \delta(\cos\theta_{\obs} - \cos\theta_{\obs}(\xvec)) \delta(\phi_{\obs} - \phi_{\obs}(\xvec)) \delta(E_{\obs} - E_{\obs}(\xvec,E_{\nu}) )
    \label{eq:Main_Delta}
\end{split}
\end{equation}
where $d\sigma_{\DM-\nu}/d\Omega_{s}$ is the differential cross section with regards to scattering angle $\theta_{s}$. The terms in the first parentheses account for the neutrino flux at position $\xvec$, while the terms in the second parentheses account for the scattering probability and eventual flux at Earth. Note that in these calculations, we consider neutrinos which scatter exactly once. Although multiple scatters are possible in theory, we typically work with cross sections small enough that multiple scatters are highly suppressed.

We will now define quantities in the integral of Eq. \ref{eq:Main_Delta} in terms of our observables. Without loss of generality, we define our coordinate system such that $\rE$ is the origin and $\rSN$ is on the x-axis. Now, using the law of cosines, we find that

\begin{equation}
    d_{x,\oplus} = \frac{2 d t + t^2}{2(d - d\cos\theta_{\obs} +t)}
    \label{eq:dex}
\end{equation}
and thus the components of $\overrightarrow{x}$ are

\begin{subequations}
    \begin{align}
        x = & d_{x,\oplus} \cos\theta_{\obs} \\
        y = & d_{x, \oplus} \sin\theta_{\obs} \sin\phi_{\obs} \\
        z = & d_{x, \oplus} \sin\theta_{\obs} \cos\phi_{\obs}.
    \end{align}
\end{subequations}

We can similarly use Eq. \ref{eq:dex} to find the scattering angle as
\begin{equation}
    \cos\theta_{s} = \frac{d^2 - (t+d-d_{x,\oplus})^2 - d_{x,\oplus}^2}{2 (t + d - d_{x,\oplus}) d_{x,\oplus}} ,\label{eq-cos-theta-scat}
\end{equation}
which is used to determine the initial neutrino energy, assuming the dark matter is initially at rest,

\begin{equation}
    E_{\nu} = \frac{E_{\obs} m_{\DM}}{m_{\DM} - E_{\obs} (1-\cos\theta_{s})}. \label{eq-Enu-calc}
\end{equation}
By defining these variables, we can now replace the integral and delta functions in Eq. \ref{eq:Main_Delta} with the Jacobian from the change of variables


\begin{equation}
\begin{split}
    &J =
    \frac{m_{\DM}^2\, t^2 (2 d + t)^2\,  \bigg[ 4 \sin^2\big(\frac{\theta_{\obs}}{2}\big)(d^2+d\,t) + t^2 \bigg]^3\,\bigg(2d\,\sin^2\big(\frac{\theta_{\obs}}{2}\big) + t\bigg)^{-4}}{8  \bigg[4 \sin^2\big(\frac{\theta_{\obs}}{2}\big)\, d^2  \bigg( 2 \sin^2\big(\frac{\theta_{\obs}}{2}\big) E_{\obs} - m_{\DM} \bigg) + (2 E_{\obs} - m_{\DM})\bigg(4 \sin^2\big(\frac{\theta_{\obs}}{2}\big)\, d\, t +  t^2\bigg)\bigg]^2}
    \end{split}
\end{equation}

\begin{equation}
\frac{d \Phi_{\obs}}{dE_{\obs} d\Omega_{\obs}} = 
J \frac{dN}{dE_{\nu}} \frac{\rho_{\DM}(\xvec)}{m_{\DM}} \frac{\exp \big(-\tau(\rSN,\xvec) - \tau(\xvec,\rE) \big)}{4 \pi d_{x,\SN}^2 d_{x,\oplus}^2} \frac{d \sigma_{\DM-\nu}}{d \Omega_{s}}
    \label{eq:Main}
\end{equation}

where $\xvec$, $E_{\nu}$, and $\theta_{s}$ are solved from the observed quantities t, $\theta_{\obs}$, $\phi_{\obs}$, and $E_{\obs}$.

By using Eqs \ref{eq:dex} and \ref{eq-cos-theta-scat}, we can calculate the time delay as

\begin{equation}
    t = d \Bigg(-1 + \cos\theta_{\obs} + \frac{\sqrt{(1-\cos^2\theta_{s}) (1-\cos^2\theta_{\obs})}}{1+\cos\theta_{s}} \Bigg) \label{eq-t-delay}.
\end{equation}

\begin{figure}
    \centering
    \includegraphics[width=0.6\linewidth]{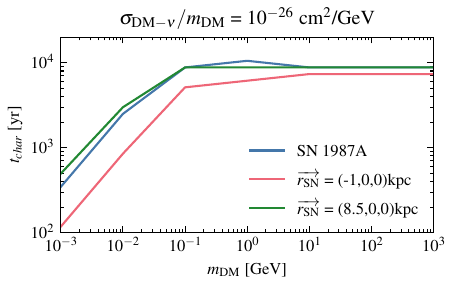}
    \caption{Characteristic neutrino delay time as a function of dark matter mass for supernova positions SN 1987A, (-1,0,0) kpc, and (8.5,0,0) kpc (note that the last point is $-\rE$). We have fixed the scattering cross section $\sigma_{\DM-\nu}/m_{\DM}$ to $10^{-26} \, \mathrm{cm^2} \, \mathrm{GeV^{-1}}$ which makes the optical depth between the supernovae and Earth is much less than 1. Furthermore, scattering is taken to be isotropic in the COM frame. We note that the value saturates at $m_{\DM} \sim 1$ GeV, at a value of $\mathcal{O}(10^{4})$ years (or $\mathcal{O}(1)$ kpc). This saturation value seems to have a slight dependence on the supernova location. \label{fig-char-times}}
\end{figure}

In the case of small scattering angles (and thus small observation angles), we obtain the $t \simeq d \theta_{s}^2/8$ result in \cite{Murase:2019xqi}. The scattering angle must follow the inequality

\begin{equation}
    \cos\theta_{s} \geq 1 - \frac{m_{\DM}}{E_{\obs}}.
\end{equation}
Furthermore, various model-dependent parameters can affect the typical scattering angle. We find when dark matter masses become comparable to or larger than the neutrino energies, these time-delays can become large (see Fig. \ref{fig-char-times}). To quantify this, we define a characteristic time $t_{char}$ such that
\begin{equation}
    \frac{d \Phi}{dE_{\rm obs}}(t_{\rm char}, E_{\rm obs}=15 \, \mathrm{MeV}) = 0.1 \frac{d\Phi}{dE_{\rm obs}}(t= 1 \, \mathrm{yr}, E_{\rm obs}=15 \, \mathrm{MeV}).
\end{equation}

We expect the rate of galactic supernovae to be $R_{SN} \sim \mathcal{O}(1)$ per century, so if $t_{char} \gg 1/R_{SN}$, we should expect a flux with contributions from multiple supernovae. One way to do this is to sample locations and times for supernova and compute Eq. \ref{eq:Main} for each one (see App. \ref{sec:Stochastic}). Alternatively, we may define a supernova rate density $n_{SN}(\overrightarrow{r})$, and compute an integral over space. Defining $\xvec$ as $\rE + r \hat{r}$ and $\rSN$ as $\rE + r \hat{r} + w \hat{w}$, our flux becomes

\begin{equation}
    \frac{d^2 \Phi}{d\Omega_{\rm MW} dE_{obs}} = \int dr dw d\Omega_{s} \frac{\rho_{\DM}(\xvec) n_{SN}(\xvec)}{4 \pi m_{\DM} } \frac{dN}{dE_{\nu}} \frac{d \sigma_{\DM-\nu}}{d\Omega_{s}} \frac{ E_{\nu}^2 \exp\big(-\tau(\rSN,\xvec) -\tau(\xvec,\rE) \big)}{E_{\obs}^2}.
    \label{eq-DMDSNB}
\end{equation}

As we are not considering one singular supernova, we have changed our observations to be relative to the Milky Way ($\theta_{MW}$ = 0 is directed towards the galactic center). We will refer to this flux as the Dark Matter Diffused Supernova Neutrino Background (DMDSNB), with the name chosen due to the similarities to the Diffuse Supernova Neutrino Background (DSNB).

\section{Models}

\subsection{Supernova Neutrinos}
As of the writing of this article, SN 1987A is the only supernova for there is confirmed the detection of neutrinos \cite{Hirata:1988ad,Bionta:1987qt}. The progenitor star was located within the Large Magellanic Cloud (LMC) with coordinates (-1.15,-43,-27.16) in kpc \cite{SN1987APos}. Considering the distribution of past galactic nuclei, we take the rate density to be

\begin{equation}
    n_{\rm SN} (\overrightarrow{r}) = A \exp \bigg( -\frac{s}{R_{d}} \bigg) \exp \bigg( - \frac{h}{H} \bigg) ,\label{eq-SN-Rate-Density}
\end{equation}
where $s$ and $h$ are radial and vertical cylindrical galactic coordinates, $R_{d} = 2.9$ kpc, $H = 95$ pc, and $A$ is a factor used to normalize the supernova rate \cite{SNLocations}. In this work, we consider the Milky Way to have a supernova rate of 1 per century.

Following \cite{Mirizzi:2015eza, Carpio:2022sml}, the supernova neutrino energies have the dependence

\begin{equation}
    \frac{dN}{dE_{\nu}} \propto \bigg( \frac{E_{\nu}}{\langle E \rangle} \bigg)^{\delta} \exp \bigg(\frac{-(\delta +1) E_{\nu}}{\langle E \rangle} \bigg),
    \label{eq-Initial-Spectrum}
\end{equation}
where we have chosen $\langle E \rangle$ = 16 MeV and $\delta$ = 2.3 and normalized so that the total energy in neutrinos is $3 \times 10^{53}$ erg. All neutrino and antineutrino flavors are assumed to have identical fluxes, and every supernova is assumed to be identical in its neutrino production. Although the average energies between different neutrino species differ, the average energy for the above single power-law fit varies around 14-18 MeV (depending on the modeling assumptions) ~\cite{Keil:2002in}. 

The DSNB, which originates from supernovae from distant galaxies, will be a background for the DMDSNB signal. The expected present-day flux of DSNB neutrinos can be estimated as an integral over redshift $z$

\begin{equation}
    \frac{d \Phi_{\rm DSNB}}{d E_{\obs}} = \int dz (1+z) \frac{dN}{dE_{\nu}}\big(E_{\obs}(1+z) \big) \frac{d \ell}{dz} R_{cc}(z), \label{eq-DSNB-Calc}
\end{equation}
where $R_{cc}(z)$ is the rate of core-collapse, and follows \cite{SNRate} 

\begin{equation}
    R_{cc}(z) \propto 
    \begin{cases}
        (1+z)^{\beta} \, \, 0<z<1 \\
        2^{\beta - \alpha} (1+z)^{\alpha} \, \, 1<z<4.5 \\
        2^{\beta-\alpha} 5.5^{\alpha - \gamma} (1+z)^{\gamma} \, \, 4.5<z<5
    \end{cases}
\end{equation}
where $\alpha = -0.26$, $\beta = 3.28$, $\gamma = -7.8$, and the supernova rate normalization $R_{cc}(0)$ is set to $1.25 \times 10^{-4}\, \text{yr}^{-1}\, \text{Mpc}^{-3}$. Furthermore, we use

\begin{equation}
    \frac{d \ell}{dz} =\bigg[ H_{0} (1+z) \sqrt{\Omega_{\Lambda}+\Omega_{M}(1+z)^{3}} \bigg]^{-1},
\end{equation}
where $H_{0}$ is the Hubble constant (here $H_0=67.4 \mathrm{km \, s^{-1} \,  Mpc^{-1}}$, $\Omega_{\Lambda} = 0.69$, and $\Omega_{\rm M} = 0.31$). Note that we treat all supernova as identical, and use this as a proof-of-concept for comparison purposes only. However, the DSNB is subject to theoretical uncertainties contained within about one order of magnitude \cite{Beacom:2010kk}, which may affect our results correspondingly.

\subsection{Dark Matter}
We describe the spatial distribution of dark matter in Galaxies with a NFW profile \cite{Navarro:1996gj}

\begin{equation}
    \rho(r) =\frac{ \rho_{0}}{(r/r_{0})(1+(r/r_{0}))^{2}}.
    \label{eq:NFW}
\end{equation}
For the Milky Way, $\rho_{0, \mathrm{MW}} = 0.184 \, \mathrm{GeV cm^{-3}}$ and $r_{0,\mathrm{MW}} = 24.42$ kpc \cite{Cirelli:2010xx}, while for the LMC, $\rho_{0, \mathrm{LMC}} = 0.31 \, \mathrm{GeV cm^{-3}}$ and $r_{0,\mathrm{LMC}} = 9.04$ kpc \cite{Siffert:2010cc}. The center of the LMC is at (-1,-40.8,-26.8) in units of kpc relative to the Milky way center (for reference, we take the Earth to be at (-8.5,0,0) kpc). For scatterings of supernova neutrinos with dark matter within the Milky Way dark matter halo, let us 
consider scattering which is isotropic in the center-of-mass frame, i.e. $d\sigma_{\DM-\nu}/d\Omega_{com} = \sigma_{\DM-\nu}/4\pi$. Boosting this into the lab-frame, we find that

\begin{equation}
    \frac{d\sigma_{\DM-\nu}}{d \Omega_{s}} = \frac{\sigma_{\DM-\nu}}{4 \pi} \frac{m_{\DM}(2 E_{\nu} + m_{\DM})}{m_{\DM}^2} \frac{E_{\obs}^2}{E_{\nu}^2}
\end{equation}
where $E_{\nu}$ and $E_{\rm obs}$ are related via Eq. \ref{eq-Enu-calc}.

\section{Observations}
\subsection{SN1987A in Neutrino Detectors}
In the neutrino burst from SN1987A, 11 events were detected at the Kamiokande-II experiment \cite{Kamiokande-II:1987idp,Hirata:1988ad} and 8 events at the Irvine-Michigan-Brookhaven (IMB) experiment \cite{Bionta:1987qt,IMB:1988suc}. Both experiments are water Cherenkov detectors, so we will consider only inverse-beta decay (IBD) processes. The IBD cross section is \cite{IBD_Cross_Sec}

\begin{equation}
    \sigma_{\rm IBD}(E_{\nu}) = \frac{7.5 \times 10^{-10} \mathrm{GeV}^{-4}}{\pi} p_{e} E_{e} = 9.36 \times 10^{-44} \mathrm{cm^2} \bigg(\frac{p_{e} E_{e}}{\mathrm{MeV^2} }\bigg),
\end{equation}
where the momentum of the outgoing positron is denoted by $p_{e}$, and its energy by $E_{e}$. The nuclear recoil is small, so $E_{e} = E_{\nu} - \Delta$, where $\Delta = 1.3 \, \mathrm{MeV}$ is the difference between the neutron and proton mass (this agrees at the percent level with \cite{Vogel:1999zy,Strumia:2003zx, Ricciardi:2022pru}). This can be converted into an event rate

\begin{equation}
    \frac{d R}{d E_{e}} = N_{H} \epsilon(E_{e}) \sigma_{\rm IBD} (E_{e}+\Delta) \frac{d \Phi(E_{e} + \Delta)}{dE_{\nu}} \label{eq-e-event-rate},
\end{equation}
where $N_{H}$ is the number of hydrogen nuclei in the detector fiducial volume ($1.4 \times 10^{32}$ for  Kamiokande and $2.2 \times 10^{32}$ for IMB), and $\epsilon(E_{e})$ is the trigger efficiency based on the positron energy. We interpolate this function from the data reported in refs.~\cite{Hirata:1988ad,Bionta:1987qt}.

We first integrate Eq. \ref{eq-e-event-rate} over $\sim$ 10 seconds (the duration of primary neutrino emission). This is equivalent to replacing $d \Phi/dE_{\nu}$ with $1/6 \times dF/dE_{\nu}$ as described using Eqs. \ref{eq-atten-fluence} and \ref{eq-Initial-Spectrum} (the $1/6$ is to account for IBD only observing anti-electron neutrinos). To obtain the total number of events, we then integrate over electron energies in the range (0.75 MeV, 100 MeV). In the case with no attenuation, this method predicts 20.98 events in Kamiokande and 13.92 events in IMB. While this is larger than the observed values, each prediction is still consistent with observation. Furthermore, this initial overestimation means an attenuation constraint will be conservative. We set a constraint when the expected number of events in Kamiokande-II (IMB) is less than 7.02 (4.66) as this is excluded by the observations at the 90$\%$ confidence-level.

\subsection{DSNB searches in Super-K}
The DSNB has not been observed with sufficiently large significance as of the writing of this paper. However, upper limits on the flux have been set by Super-Kamiokande (Super-K), a 22.5 kiloton-size water Cherenkov detector designed to study atmospheric neutrinos. In particular, Run IV, which incorporated 2970 days of data-taking from 2008 to 2018, has set some of the strongest limits on the flux. In ref.~\cite{SKRunIVDSNB}, the energy window of (9.3 MeV, 31.3 MeV) is investigated in bins 2 MeV in width, with upper limits on the flux provided for each of these bins. \footnote{Specifically, the data in their Table V and Figure 25 are utilized in our analysis.}

We currently ignore angular information about the neutrino flux, meaning that neutrinos either from a single source or the DMDSNB (calculated using Eq. \ref{eq:Main} and Eq. \ref{eq-DMDSNB} respectively) will contribute to searches for the DSNB. We exclude a parameter point if it leads to a delayed neutrino flux which supersedes the flux upper limits from ref.~\cite{SKRunIVDSNB} in any of the 2 MeV bins.

\section{Results}
Although we compute cross section constraints for the DMDSNB for $10 \, \mathrm{keV} < m_{\DM} < 100 \, \mathrm{TeV}$, we only consider the constraints valid for $m_{\DM} \gtrsim 100$ MeV. For smaller masses, the DMDSNB flux is not expected to be constant with the neutrino energy (see the discussion in App. \ref{sec:Stochastic}). On the other hand, no restrictions are placed on the dark matter masses for SN 1987A attenuation or time-delayed neutrino results.

\begin{figure}[t]
    \subfloat{
      \includegraphics[width=0.45\linewidth]{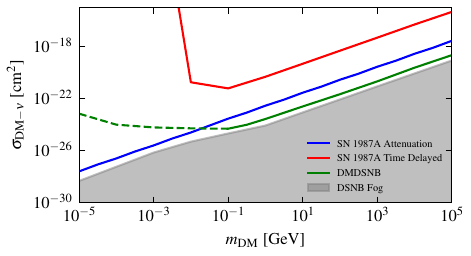}
    }
    \hfill
    \subfloat{
      \includegraphics[width=0.45\textwidth]{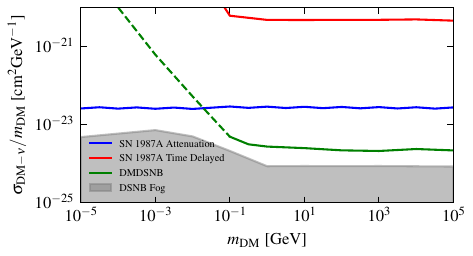}
    }
    \caption{\textbf{Left:} Excluded value of $\sigma_{\DM-\nu}$ as a function of $m_{\DM}$ from SN 1987A attenuation, SN 1987A time-delayed neutrinos at Super-K, and the DMDSNB at Super-K. Note that for dark matter masses below 100 MeV, the DMDSNB exclusions are dotted, as we do not expect the fluxes to be constant in time (see App. \ref{sec:Stochastic}).  \textbf{Right:} The same exclusions expressed as a $\sigma_{\DM-\nu}/m_{\DM}$ ratio.\label{fig-cross-sec-results}}
\end{figure}

We can see the exclusions on $\sigma_{\DM-\nu}$ and on $\sigma_{\DM-\nu}/m_{\DM}$ in Fig. \ref{fig-cross-sec-results}.  The attenuation constraint from SN1987A is always stronger than the constraint from time-delayed neutrinos solely from SN1987A.  This is to be expected, as the supernova occurred at a relatively large distance from Earth. The DMDSNB provides the strongest constraint for masses above $m_{\DM}>100$ MeV. When considering $\sigma_{\DM-\nu}/m_{\DM}$, we see that the constraints become essentially constant for $m_{\DM} > 1$ GeV, saturating to $2.7 \times 10^{-23}\,\mathrm{cm^2\, GeV^{-1}}$, $4.8\times 10^{-21}\mathrm{cm^2\,GeV^{-1}}$, and $2.4 \times 10^{-24}\,\mathrm{cm^2\, GeV^{-1}}$ for SN1987A Attenuation, SN1987A time-delayed neutrinos, and the DMDSNB respectively. The DMDSNB constraint changes only by an $\mathcal{O}(1)$ factor for $\langle E \rangle$ between 10 and 20 MeV.

If the DMDSNB flux is expected to exceed the DSNB flux, then constraints can be made simply through an upper limit on the neutrino flux. For smaller $\sigma_{\DM-\nu}$ (and thus smaller DMDSNB fluxes), this effect can be observed through shifts in the neutrino spectrum (see Fig. \ref{fig-DSNB-DMDSNB-Compare}). We will define the $\sigma_{\DM-\nu}$ at which the DMDSNB flux is below the predicted DSNB flux for the whole range 10-30 MeV as the ``DSNB fog" (as an analog to the ``neutrino fog" or ``neutrino floor" for dark matter experiments \cite{NeutrinoFloor,OHare:2021utq, Wyenberg:2018eyv, Herrera:2023xun,Maity:2024vkj, Carew:2023qrj}). We show the result for the DSNB fog in Fig. \ref{fig-cross-sec-results}. While our computation of the DSNB fog is informative and this point was previously absent in supernova-related dark matter studies, we want to end with two caveats. First, we are still using our simplified model for the DSNB in which all supernova are identical; however, theoretical predictions range over over an order of magnitude \cite{SKRunIVDSNB}. If a more detailed supernova model is used to make DSNB predictions, that same model should be used to predict the DMDSNB. Second, reaching the DSNB fog does not mean constraints can no longer be placed on this scenario, only that the method of placing these constraints will change and become dependent on the true DSNB flux.

\section{Comparison to complementary bounds}

We contextualize the strength of our constraints compared to other probes of dark matter-neutrino scatterings here. Strong cosmological bounds on dark matter-neutrino interactions have been derived, relying on modifications of the matter power spectrum at small scales, \textit{e.g} \cite{Crumrine:2024sdn, Akita:2023yga, Brax:2023tvn}. These bounds apply on comoving wavenumbers of $k \sim 10-100 h$ Mpc$^{-1}$. We can find the redshift of these scales and the corresponding temperature of the Universe, which inform us about the energy scale at which these bounds apply. Perturbations enter the horizon when their physical scale matches the Hubble radius $k=aH(a)$. In the radiation-dominated era we have $k=H_0 \sqrt{\Omega_r} a^{-1}$, and, relating the redshift to the scale factor as $(1+z)=a^{-1}$ we find (using $H_0=100 h \mathrm{~km} / \mathrm{s} / \mathrm{Mpc}, \Omega_r \sim 9 \times 10^{-5}$, $T_{\rm CMB}=2.35 \times 10^{-4} \mathrm{eV}$, and $h\sim 0.7$) that the energy scale becomes $T \sim T_{\rm CMB} k/(H_0 \sqrt{\Omega_r}) \sim 0.73-7.3 \, \mathrm{keV}$.
\begin{figure}
\centering
      \includegraphics[width=0.49\linewidth]{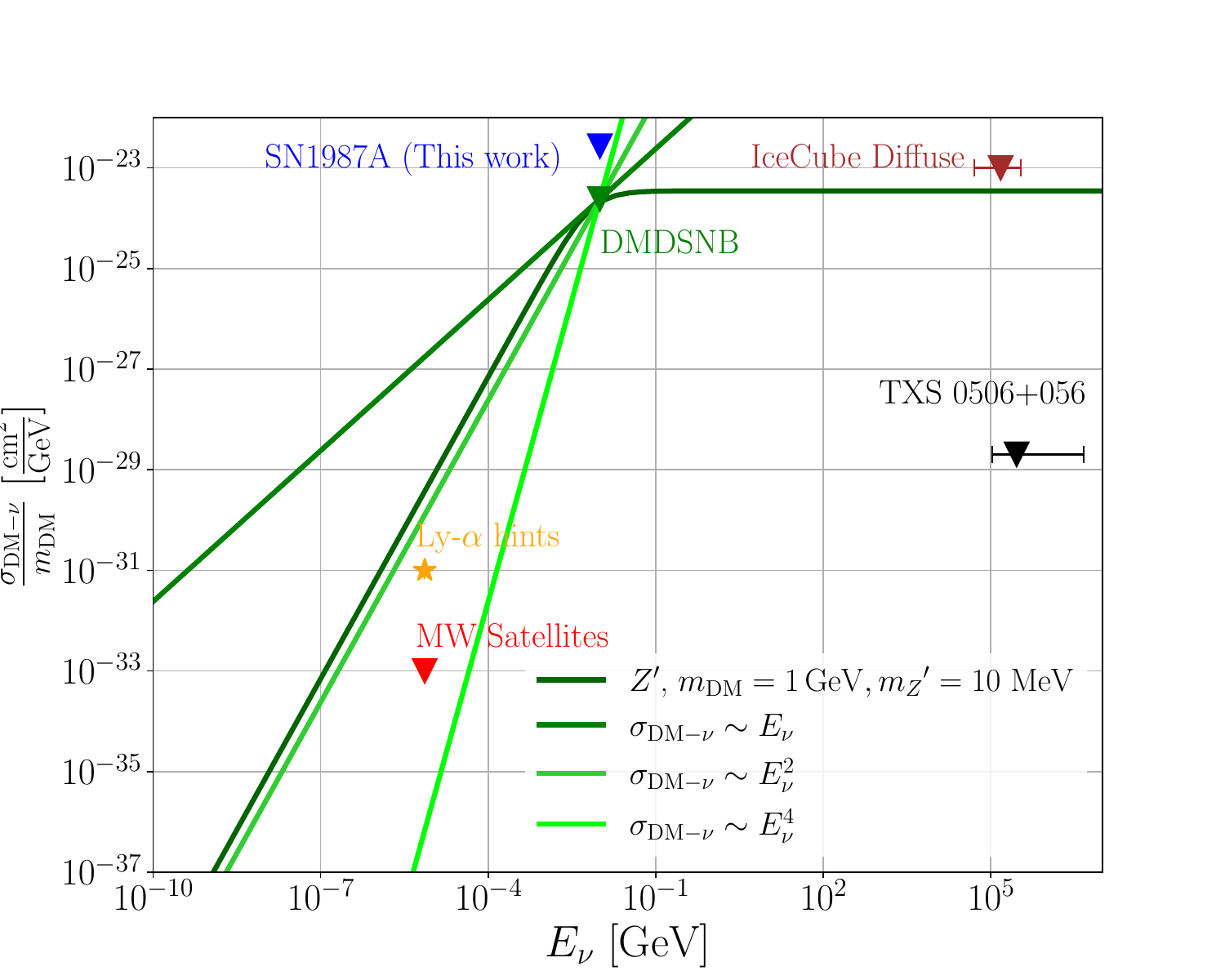}
    \includegraphics[width=0.49\linewidth]{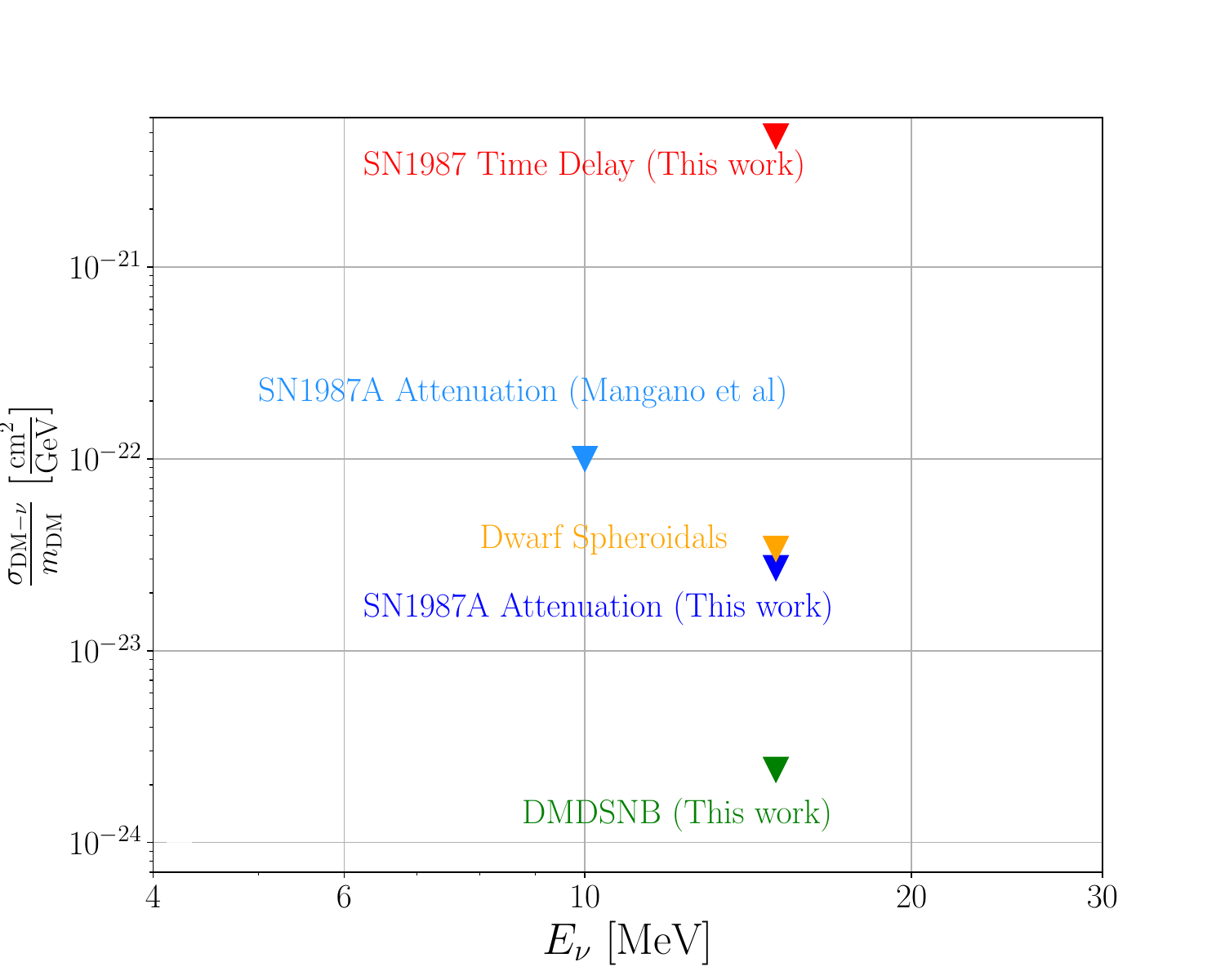}
    \caption{\textbf{Left}: Summary of bounds on dark matter-neutrino scatterings across neutrino energy. The bounds derived in this work from SN1987A and from the Dark Matter Diffused Supernova Neutrino Background (DMDSNB) are shown as blue and green triangles, respectively. Furthermore, we show in green lines the strength of our limits at different neutrino energies, for different phenomenological dependences of the cross section with neutrino energy. Our constraints are compared to cosmological hints and bounds (in orange and red) \cite{Crumrine:2024sdn, Hooper:2021rjc}, and to bounds from high-energy neutrinos (in brown and black) \cite{Arguelles:2017atb, GonzaloTXSAtten}. \textbf{Right}: Summary of bounds on the dark matter-neutrino scattering cross section from supernovae. Our bounds on attenuation effects from SN1987A (blue), time delays (red), and the DMDSNB (green)  are compared with previous bounds on attenuation effects in SN1987A (light blue) where they did not use an NFW profile nor did they consider the dark matter enhancement in the LMC \cite{Mangano:2006mp}, and with bounds from the impact of dark matter-neutrino scatterings on dwarf spheroidals \cite{Heston:2024ljf}. \label{fig-Constraint-Compare}}
\end{figure}
We show in the left panel of Fig. \ref{fig-Constraint-Compare} the strongest cosmological bound as a red arrow, arising from Milky Way satellites \cite{Crumrine:2024sdn}, and the hinted value of dark matter-neutrino interactions from Ly-$\alpha$ \cite{Hooper:2021rjc} at somewhat larger cross sections. For a proper comparison of astrophysical and cosmological bounds, we scale the strength of our bounds across neutrino energy for different phenomenological dependences of the dark matter-neutrino scattering cross section. The lighter green line illustrates the strength of our constraints for a cross section scaling as $\sigma_{\DM-\nu} \sim E_{\nu}^4$, which is expected for higher dimensional ($n>6$) operators. We also show constraints for a cross section scaling as $\sigma_{\DM-\nu} \sim E_{\nu}^2$, which arises in $\textit{e.g}$ scenarios where dark matter is a complex scalar that interacts with neutrinos via a fermion portal, with a Yukawa term as $\sim \Phi \bar{\chi} \nu_{L}$, with $\Phi$ being the dark matter field and $\chi$ the fermion mediator. In dark green lines, we show a linear scaling of the cross section.
For the darkest line, we consider a vector mediator and fix the dark matter mass to $m_{\rm DM}=1$ GeV, and the mediator mass to $m_{Z^{\prime}}= 10$ MeV. In the low-energy limit, the cross section scales with energy as
\begin{align}
   \sigma_{\rm DM-\nu} \simeq \, &\frac{g_{\text{DM}}^2\, g_\nu^2}{4\, \pi}\dfrac{E_\nu ^2}{M_{Z'}^4},
\end{align}
and, in the high-energy limit, the cross section becomes flat
\begin{align}
    \sigma_{\rm DM-\nu} &\simeq \frac{g_{\rm DM}^2 g_\nu^2}{8\, \pi}\dfrac{1}{M_{Z^\prime}^2}.
\end{align}
It can be appreciated that our constraints, when extrapolated to low energies, become comparable to cosmological bounds in models with a quadratic scaling of the cross section, and would become stronger for high-dimensional operators with a quartic scaling. In addition, the bounds derived in this work from supernovae dark matter-neutrino interactions, comparable bounds have been derived from core-collapse supernovae within Milky Way spheroidal galaxies \cite{Heston:2024ljf}. We also show bounds from DM-neutrino interactions attenuating the flux of SN1987A from \cite{Mangano:2006mp}. These are somewhat weaker than out derived limits from the same consideration, since their modelling of the galactic dark matter density profile scales with $r^{-2}$, while we use an NFW profile scaling as $r^{-1}$, thus the column density along the ling of sight of the LMC is larger in our case. We further show constraints arising from high-energy neutrino observations in IceCube; in brown we show constraints from the attenuation of the astrophysical diffuse neutrino flux in the Milky Way halo \cite{Arguelles:2019ouk}, and in black we show constraints arising from the attenuation in the blazar TXS 0506+056 \cite{GonzaloTXSAtten}. Our constraints improve upon the IceCube ones from the diffuse astrophysical flux in models where the cross section is flat or decreases with energy at high energies, but are orders of magnitude weaker than those from TXS 0506+056. Those constraints suffer from uncertainties on the dark matter distribution at TXS 0506+056, and could be relaxed if the dark matter density is significantly suppressed in the vicinity of the supermassive black hole compared to theoretical expectations. In the right panel of Fig. \ref{fig-Constraint-Compare}, we show a summary of constraints at MeV energies from this work and from previous related studies \cite{Mangano:2006mp,Heston:2024ljf}.

\section{Outlook and Conclusions}
\begin{figure}[t]
\centering
    \includegraphics[width = 0.49\textwidth]{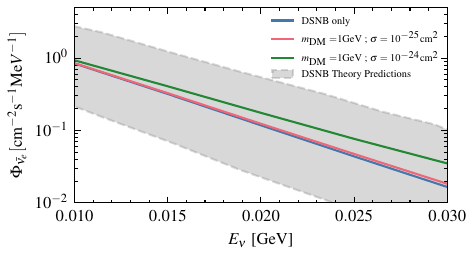}
    \includegraphics[width = 0.49 \textwidth]{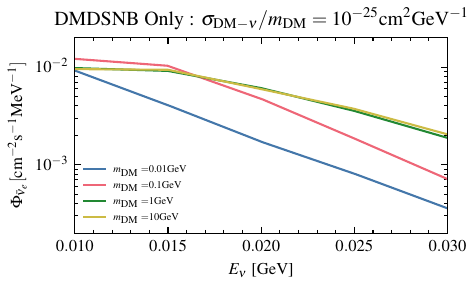}
    \caption{Flux vs observed neutrino energy for different scenarios. \textbf{Left: }The blue line considers only the DSNB contribution calculated using Eq. \ref{eq-DSNB-Calc}. The red (green) lines show the combined flux from the DSNB and DMDSNB with dark matter of $m_{\DM} =$1 GeV and $\sigma_{\DM-\nu} = 10^{-25} \,(10^{-24})\, \mathrm{cm^2}$. We see that especially at higher energies, the expected fluxes of cases with and without the DMDSNB differ. The grey shaded region indicates the range of theory predictions (see \cite{SKRunIVDSNB} and references therein). \textbf{Right:} The neutrino flux solely from the DMDSNB for dark matter of different masses. We fix the ratio $\sigma_{\DM-\nu}/m_{\DM} = 10^{-25}$ cm$^2$ GeV$^{-1}$. \label{fig-DSNB-DMDSNB-Compare}}
\end{figure}

Our limit of $\sigma_{\rm DM-\nu} \lesssim 2.4 \times 10^{-24} \,\mathrm{cm^2\, GeV^{-1}}$, derived from the non-observation of the DMDSNB in Super-Kamiokande, is the strongest limit to date at MeV neutrino energies and can be further improved. As stronger upper bounds are set on the DSNB flux, these same bounds can be applied to the DMDSNB and strengthen our limit on $\sigma_{\DM-\nu}$. Thus we can expect our limits to improve even if a galactic supernova does not occur in the near-future. Furthermore, an observation of the DSNB does not prevent improving constraints on the DMDSNB. First, the expected energy distribution of the DMDSNB is different from the DSNB (see Fig. \ref{fig-DSNB-DMDSNB-Compare}). This is to be expected, as the DSNB evolves with redshift and must have a peak at lower energies, while time-delayed neutrinos have nearly the same energy distribution as a galactic supernova provided $m_{\DM} \gg E_{\nu}$ (see Eq. \ref{eq-Enu-calc}). Thus, even when the DSNB is discovered, we can distinguish a DMDSNB contribution from an excess in the higher energy bins. The exact spectrum may even provide information on the dark matter mass (see the right panel of Fig. \ref{fig-DSNB-DMDSNB-Compare}.

\begin{figure}[t]
\centering
    \includegraphics[width = 0.7\textwidth]{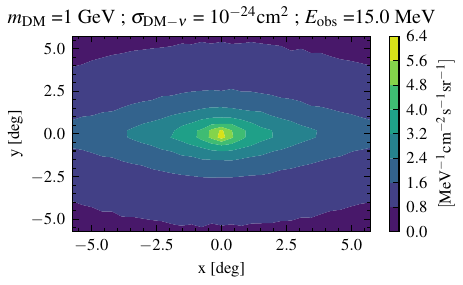}
    \caption{Differential flux of the DMDSNB as observed from Earth. The point of $x=y=0^{\circ}$ is directed towards the galactic center. The x-y asymmetry is from the supernova distribution extending further in the radial direction than the horizontal, and the strong focus towards the central point due to the high dark matter density near the galactic center. \label{fig-angles}}
\end{figure}

We should also note that while the DSNB is isotropic, the DMDSNB is strongly focused toward the galactic center due to the higher density of dark matter and increased supernova rate (see Fig \ref{fig-angles}). This feature may be more difficult to detect with IBD since these events are nearly isotropic, and angular information would likely need to be extrapolated from $\nu-e$ scatterings. 

This work could be extended in many ways. First, substantial dark matter-neutrino scattering would also alter the expectations of the DSNB, as scattering can occur in the supernova host galaxy, in the intergalactic medium, and in the Milky Way as the neutrinos approach Earth. The contribution from the intergalactic medium is expected to be comparable to the contribution from the Milky Way, but the contribution from the distant host galaxy may be significantly larger depending on the spatial distribution of supernovae. In some models the supernova distribution peaks at the galactic center \cite{Verberne:2021tse, Ranasinghe:2022ntj}, where the integrated effect with the dark matter density could be significant. In the case DMDSNB flux exceeding the DSNB flux and poor directional resolution, the presence of DMDSNB could also lead to degeneracies between different DSNB model predictions with a harder spectrum versus true DSNB flux with a softer spectrum. We leave detailed calculations of these scenarios for a future study.

Additionally, although we use supernova and neutrinos in this work, the idea of an astrophysical flux diffused by galactic dark matter can be applied to other astrophysical sources and feebly interacting particles.  This can be applied as long as the induced time delay is substantially longer than the time elapsed between these events.

\section*{Acknowledgments}
We would like to thank Andrew Caruso and Sarah Healy for helpful discussions on DSNB observations and stellar distributions. We also thank Damiano F.G. Fiorillo for useful discussions on SN1987A. The work of GC, RAG, GH and IS is supported in part by the U.S. Department of Energy under the award number DE-SC0020262. The work of GC is also supported by the NSF Award Number 2309973. RAG is partially supported by the World Premier International Research Center Initiative (WPI), MEXT, Japan. RAG is grateful to QUP for hospitality during his visit. 
This material is based upon work supported by the U.S. Department of Energy, Office of Science, Office of Workforce Development for
Teachers and Scientists, Office of Science Graduate Student Research (SCGSR) program. The SCGSR program is administered by the
Oak Ridge Institute for Science and Education (ORISE) for the DOE. ORISE is managed by ORAU under contract number
DESC0014664. All opinions expressed in this paper are the author’s and do not necessarily reflect the policies and views of DOE,
ORAU, or ORISE. RAG is a recipient of the SCGSR Award.

\appendix
\section{Stochastic Nature of DMDSNB \label{sec:Stochastic}}
\begin{figure}[H]
\centering
    \includegraphics[width = 0.7 \textwidth]{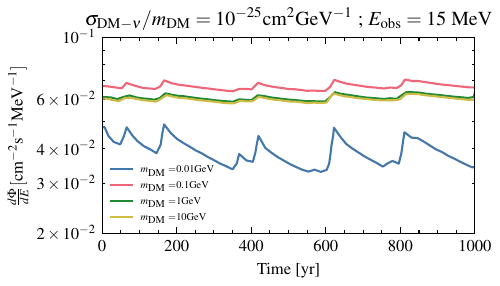}
    \caption{All flavor neutrino flux vs time for the DMDSNB considering different dark matter masses with $\sigma_{\DM-\nu}/m_{\DM}$ fixed to $10^{-25} \mathrm{cm}^{2} \mathrm{GeV}^{-1}$. The fluxes are computed from simulating supernova with a spatial distribution described by Eq. \ref{eq-SN-Rate-Density} and an average rate of 1 per century. The same supernova times and positions are used for each mass point. \label{fig-Flux-vs-Time-Stochastic}}
\end{figure}

In the case where typical time delays are $\mathcal{O}(10^{3}$ years), we are able to average over many supernovae and obtain a flux that is essentially constant in time. However, for shorter time delays, the stochastic nature of supernovae appears in the DMDSNB flux. As can be seen in Fig. \ref{fig-Flux-vs-Time-Stochastic}, for dark matter masses above $m_{\rm DM} \gtrsim 100$ MeV, the flux stays nearly constant, while for $m_{\DM} = 10$ MeV, there are variations in the flux of $\mathcal{O}(1)$. For this reason, we do not set constraints below $m_{\rm DM} < 100$ MeV, as the true behavior cannot be described via the average flux.

\section{Impact of a Milky Way Dark Matter Spike}
In this work, we have considered a simple NFW profile for our dark matter. This prescription is unlikely to hold in the region of gravitational influence of the central black hole, where the supernova density may be enhanced compared to other regions of the Galaxy \cite{Verberne:2021tse}. It is expected that a central dark matter spike is formed around supermassive black holes (SMBHs) \cite{Quinlan:1994ed,Gondolo:1999ef}. In this appendix, we compute the resulting neutrino flux from dark matter scatterings off supernova neutrinos scatterings within the spike around Sagittarius $A^{*}$. We will let $\rho_{\rm sp}(r) \sim r^{-\gamma}$, $ 1\leq \gamma \leq 3$ out to an extent $R_{\max}$ with the requirements

\begin{equation}
    M_{\rm BH} \simeq 4 \pi \int_{0}^{R_{\max}} r^2 \rho_{\rm sp}(r) dr,
\end{equation}

\begin{equation}
    \rho_{\rm sp}(R_{\max}) = \rho_{\rm NFW}(R_{\max}).
\end{equation}

\begin{figure}
    \includegraphics[width = 0.45 \textwidth]{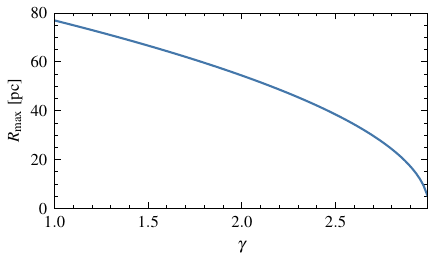}
    \includegraphics[width = 0.5 \textwidth]{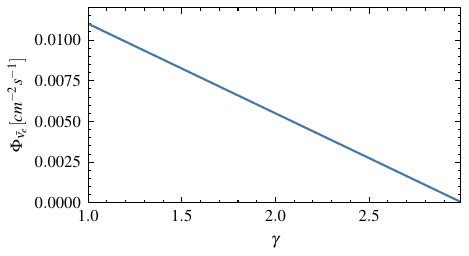}
    \caption{\textbf{Left:} Radius of the dark matter spike in the Milky Way as a function of the power-law index $\gamma$. \textbf{Right:} Flux of scattered neutrinos at Earth as a function of $\gamma$. Method of determining the flux is described in the text. \label{fig-Spike}}
\end{figure}

The first condition is motivated by the kinematics of S2-stars around Saggitarius A$^{*}$, which does not allow for an enclosed spike mass much larger than the black hole mass \cite{Lacroix:2018zmg}. We show the resulting extent of the spike as a function of $\gamma$ in the left panel of Fig. \ref{fig-Spike}. Next, to estimate the resulting flux, we will assume every neutrino which enters the spike undergoes scatterings with dark matter particles, and the resulting scattered flux is isotropic. Let $ r_{c}$ be the characteristic distance of a supernova from the galactic center, $\Gamma_{SN}$ be the supernova rate, and $N_{\nu}$ be the number of neutrinos produced per supernova. Then

\begin{equation}
    \Phi_{\bar \nu_{e}} \simeq \frac{\Gamma_{\rm SN}}{6} \frac{N_{\nu}}{4 \pi r_{\oplus}^2} \bigg(\frac{R_{\max}}{2 r_c} \bigg)^2.
\end{equation}

Taking $r_{c}$=2.9 kpc and $\Gamma_{\rm SN}=0.01\mathrm{yr}^{-1}$ we obtain the fluxes see in the right panel of Fig. \ref{fig-Spike}. The resulting fluxes are smaller than what is currently constrained by Super-K, but this only holds as long as the supernova rate $\Gamma_{\rm SN}$ is not significantly enhanced towards the galactic center, a possibility we cannot preclude. Furthermore, future work considering time-delayed neutrinos in more detail might allow us to find distinctive spatial or temporal features in the neutrino flux due to the presence of a dark matter spike.

\section{Dependence on Average Supernova Energy}

\begin{figure}
\centering
    \includegraphics[width = 0.7 \textwidth]{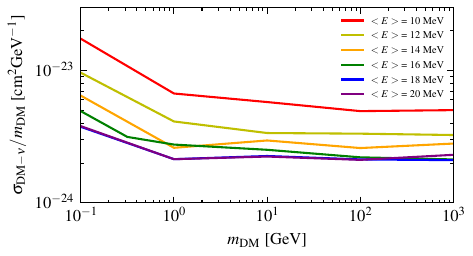}
    \caption{ Exclusions on the cross section vs mass due to the DMDSNB with different average energies. We still fix $\delta = 2.3$, the total neutrino luminosity per supernova is $3 \times 10^{53} $ erg, and there is an average rate of one galactic supernova per century. } \label{fig-diff-avg-energy}
\end{figure}

For all DMDSNB results obtained in this work, we have assumed the supernova spectrum followed Eq. \ref{eq-Initial-Spectrum} with $\langle E \rangle =$16 MeV. While still taking this simple expression for the supernova spectrum, we show how DMDSNB constraints change with $\langle E \rangle$ in Fig. \ref{fig-diff-avg-energy}. For average energies between 10 and 20 MeV, the bounds change only by $\mathcal{O}(1)$, with higher energies providing stronger constraints. We note that the power-law fits are only expected to have average energies in the range 14-18 MeV ~\cite{Keil:2002in} so this shows our bounds are especially robust. This is to be expected, as the SK limits on the neutrino flux become stronger with increasing energy \cite{SKRunIVDSNB}.

\bibliographystyle{bibi}
\bibliography{main}

\end{document}